\documentclass[sigconf,natbib=true]{acmart}
\usepackage{hyperref}
\usepackage{multirow}
\AtBeginDocument{%
  }
\copyrightyear{2025}
\acmYear{2025}
\setcopyright{cc}
\setcctype{by}
\acmConference[SIGIR '25]{Proceedings of the 48th International ACM SIGIR Conference on Research and Development in Information Retrieval}{July 13--18, 2025}{Padua, Italy}
\acmBooktitle{Proceedings of the 48th International ACM SIGIR Conference on Research and Development in Information Retrieval (SIGIR '25), July 13--18, 2025, Padua, Italy}\acmDOI{10.1145/3726302.3730216}
\acmISBN{979-8-4007-1592-1/2025/07}

\begin{document}

\title{Investigating Task Arithmetic for Zero-Shot Information Retrieval}

\author{Marco Braga}
\email{m.braga@campus.unimib.it}
\affiliation{%
  \institution{Department of Informatics, Systems and Communication - DISCo, University of Milano-Bicocca}
  \city{Milano}
  \country{Italy}
}
\affiliation{%
  \institution{DAUIN Dipartimento di Automatica e Informatica, Politecnico di Torino}
  \city{Turin}
  \country{Italy}
}
\author{Pranav Kasela}
\email{pranav.kasela@unimib.it}
\affiliation{%
  \institution{Department of Informatics, Systems and Communication - DISCo, University of Milano-Bicocca}
  \city{Milano}
  \country{Italy}
}

\author{Alessandro Raganato}
\email{alessandro.raganato@unimib.it}
\affiliation{%
  \institution{Department of Informatics, Systems and Communication - DISCo, University of Milano-Bicocca}
  \city{Milano}
  \country{Italy}
}

\author{Gabriella Pasi}
\email{gabriella.pasi@unimib.it}
\affiliation{%
  \institution{Department of Informatics, Systems and Communication - DISCo, University of Milano-Bicocca}
  \city{Milano}
  \country{Italy}
}


\begin{abstract} 

Large Language Models (LLMs) have shown impressive zero-shot performance across a variety of Natural Language Processing tasks, including document re-ranking. However, their effectiveness degrades on unseen tasks and domains, largely due to shifts in vocabulary and word distributions. In this paper, we investigate Task Arithmetic, a technique that combines the weights of LLMs pre-trained on different tasks or domains via simple mathematical operations, such as addition or subtraction, to adapt retrieval models without requiring additional fine-tuning. Our method is able to synthesize diverse tasks and domain knowledge into a single model, enabling effective zero-shot adaptation in different retrieval contexts. Extensive experiments on publicly available scientific, biomedical, and multilingual datasets show that our method improves state-of-the-art re-ranking performance by up to 18\% in NDCG@10 and 15\% in P@10. In addition to these empirical gains, our analysis provides insights into the strengths and limitations of Task Arithmetic as a practical strategy for zero-shot learning and model adaptation. We make our code publicly available at \url{https://github.com/DetectiveMB/Task-Arithmetic-for-ZS-IR}. 

\end{abstract}





\maketitle

\section{Introduction}

Large Language Models (LLMs) have achieved state-of-the-art performance in a wide range of tasks in several research fields \cite{Survey_llm, rogers2024position}, including Information Retrieval (IR) \cite{zhu2023large}. 
By learning representations from massive unlabeled corpora, LLMs can be employed in document re-ranking \cite{nogueira-etal-2020-document,qin-etal-2024-large}, query expansion \cite{chen-etal-2024-analyze,Li_query, Raganato_queryexp}, and synthetic data generation \cite{almeida-matos-2024-exploring, braga2024synthetic}. Notably, these models often excel in zero-shot scenarios \cite{NEURIPS2020_1457c0d6,agrawal-etal-2022-large,wang2024zero}, effectively handling unseen tasks or domains without additional supervised fine-tuning. This zero-shot capability has driven their widespread use in document re-ranking \cite{Setwise_zeroshot,Tip_Borges,lin2022pretrained}, enabling robust retrieval even in the absence of domain-specific training data.
Despite these advantages, domain mismatch remains a critical challenge that can significantly 
hamper the effectiveness of the model
\cite{Wang_domain}. The BEIR benchmark \cite{thakur2021beir,BEIR_res}, which spans diverse tasks and domains, provides a heterogeneous framework for evaluating zero-shot IR performance. A common strategy is to pre-train a model on a large-scale IR dataset (e.g., MS-MARCO \cite{MSMARCO}) and then apply it zero-shot to unseen domains \cite{BEIR_res}. Although this approach often achieves strong performance, developing a single LLM that robustly generalizes to every domain remains 
very
challenging \cite{thakur2021beir, kasela2024desire}, as domain-specific adaptation typically 
requires large
labeled 
datasets
and 
considerable computational resources \cite{xia2024understanding}.
To mitigate these costs, Parameter-Efficient Fine-Tuning (PEFT) approaches have been proposed \cite{houlsby2019parameter,hulora, braga-etal-2024-adakron}, which adapt a small subset of parameters while leaving most of the model frozen. Although PEFT significantly reduces the scale of updates and can be effective under limited data, labeled instances are still needed for each target domain, 
which makes it less suitable for frequent domain shifts or truly zero-shot scenarios.
Meanwhile, the proliferation of publicly available LLMs, fine-tuned for diverse tasks, domains, and languages, on open-source platforms such as HuggingFace \cite{jain2022hugging}, offers a new opportunity to reuse existing models rather than training new ones from scratch. In this context, Task Arithmetic \cite{ilharcoediting} emerges as a promising method. In fact, by merging the parameters of two or more fine-tuned LLMs through a simple addition or subtraction, Task Arithmetic transfers domain knowledge into a target model without further gradient-based optimization. More specifically, a Task Vector is defined as the difference between the parameters of a domain-fine-tuned model and its original pre-trained version, allowing domain-specific information 
to be injected or removed in parameter space with minimal overhead.

\begin{figure}
    \centering
    \includegraphics[width=0.6\linewidth]{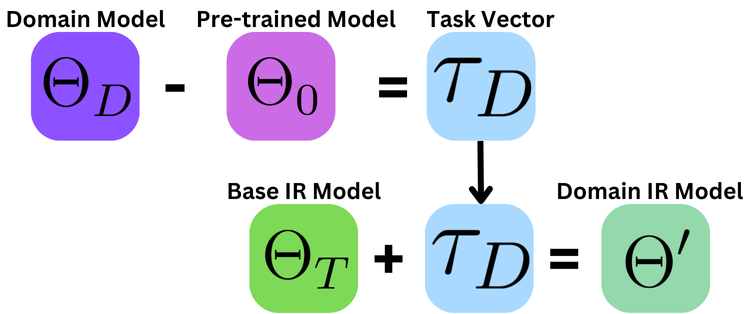}
    \caption{Proposed approach: Given a pre-trained LLM $\Theta_0$ and its domain-finetuned version $\Theta_D$, we compute the Task Vector $\tau_D$ as their parameter difference. To build a domain-specific IR model $\Theta'$, we add $\tau_D$ to an IR-finetuned model $\Theta_T$.}
    \Description{Our proposed approach: given a pre-trained Large Language Model $\Theta_0$ and its finetuned version on a specific domain $\Theta_D$, we obtain the Task Vector $\tau_D$ as the difference between the parameters of these models. To define a domain-specific IR model $\Theta'$, we add $\tau_D$ to a model $\Theta_T$ already fine-tuned for the IR task.}
    \label{fig:task-vector-img}
\end{figure}

In this work, we investigate how Task Arithmetic can be applied to Information Retrieval, emphasizing domain and language transfer under zero-shot conditions. As shown in Figure~\ref{fig:task-vector-img}, we begin with a pre-trained LLM and leverage domain-specialized vectors derived from fine-tuned models, combining them with an IR-focused baseline to produce a new domain-aware 
model. Our extensive evaluation covers eight publicly available datasets spanning scientific, biomedical, and multilingual tasks. We experiment with six LLMs with parameter counts ranging from 66 million to 7 billion, including encoder-only, encoder-decoder, and decoder-only architectures. Inasmuch as all specialized models are publicly available, our method remains reproducible, computationally efficient, and 
it does not require
additional training 
thus
minimizing carbon footprint.
Our results show that Task Arithmetic consistently improves upon strong IR baselines, with gains of up to 18\% in NDCG@10 and 15\% in P@10, highlighting the practical value of reusing existing domain-trained models and 
underlying how
Task Arithmetic 
is
a lightweight 
but
powerful strategy for zero-shot domain and language adaptation in IR.

\section{Methodology}
\label{Metodo}

In this Section, we first present related work on 
model merging 
and introduce Task Arithmetic (Section~\ref{task_ar}). 
We then explain
how we adapt this framework for zero-shot IR (Section~\ref{sec:tzir}). 


\subsection{Related Work and Motivation}
\label{task_ar}

Adapting LLMs to specialized domains typically requires resource-intensive fine-tuning. To address frequent domain shifts or label-scarce settings, recent studies explore weight interpolation and model merging \cite{wortsman2022model, matena2022merging}, leveraging the observation that models fine-tuned on related tasks may reside in compatible regions of parameter space \cite{ainsworth2023git, choshen2022fusing}. This allows arithmetic operations, such as averaging parameters, to preserve or even boost performance \cite{li2022branchtrainmerge, Frankle2020The, izmailov2018averaging}. 
Among these methods, Task Arithmetic \cite{ilharcoediting} is particularly appealing 
for the fact that it requires no training.
Each domain- or task-specific fine-tuning run is represented as a Task Vector, defined by the difference between the fine-tuned model’s parameters and those of the pre-trained model; adding or subtracting these vectors transfers knowledge without further updates \cite{daheim-etal-2024-elastic, chronopoulou-etal-2024-language, parovic-etal-2024-investigating, huang-etal-2024-chat, bhardwaj-etal-2024-language}. Unlike approaches requiring adapters \cite{houlsby2019parameter, Kasela_braga_sepqa, braga2023personalization, braga_sigir} or gating modules \cite{mao2022unipelt, sukhbaatar2024branch, kasela2024desire}, Task Arithmetic retains the original network architecture and remains cost-effective. 
In the context of zero-shot Information Retrieval (IR), Task Arithmetic remains relatively underexplored. Common approaches often rely on domain-specific adapters or fine-tuning \cite{yadav2023ties, yu2024language,  bassani2022multi, zhan2022disentangled, artetxe-etal-2020-cross}, which are effective but still require training data and overhead. In contrast, Task Arithmetic simply reuses existing domain-fine-tuned LLMs, readily available on open-source platforms, and integrates them plug-and-play into an IR model. This approach is particularly appealing given the recent proliferation of LLMs fine-tuned for diverse domains and languages; by merging their specialized capabilities, we can construct a domain-aware IR model with minimal computational cost.

\begin{table*}[htp]
\resizebox{\linewidth}{!}{
\begin{tabular}{cc|cccc|cccc|cccc|cccc|}

\multicolumn{2}{c|}{\textbf{Model}} & \multicolumn{4}{c|}{\textbf{SciFact}}   & \multicolumn{4}{c|}{\textbf{NFCorpus}} & \multicolumn{4}{c|}{\textbf{SCIDOCS}} & \multicolumn{4}{c}{\textbf{TREC-COVID}} \\ \midrule

\multicolumn{1}{|c|}{\textbf{Re-ranker}} & \multicolumn{1}{c|}{Variant}  & \small{P@10} & \small{NDCG@3} & \small{NDCG@10} & \small{MAP@100}  & \small{P@10} & \small{NDCG@3} & \small{NDCG@10}  & \small{MAP@100}  & \small{P@10} & \small{NDCG@3} & \small{NDCG@10}  & \small{MAP@100}  & \small{P@10} & \small{NDCG@3} & \small{NDCG@10} & \small{MAP@100} \\ \midrule

\multicolumn{2}{|c|}{BM25} & .091 & .637 & .691 & .649 & .247 & .404 & .343 & .154 & .086 & .156 & .165 & .112 & .734  &.764 & .688 & .085  \\ \midrule

\multicolumn{1}{|c|}{\multirow{4}{*}{\texttt{Llama-2-7B}}} & $\Theta_0$: Pre-trained  & .\textbf{099} & .701 & .748 & .702 & .260 & .431 & .363 & .165 & .\textbf{099} & .179 & .\textbf{191} & .\textbf{130} & .838 & .845 & .782 & .095 \\  

\multicolumn{1}{|c|}{} & $\Theta_D$: Domain-specific (MedTuned)
& .097 & .699 & .742 & .701 & .250 & .410 & .349 & .158 & .095 & .175 & .184 & .124 & .812 & .810 & .761 & .093 \\ 

\multicolumn{1}{|c|}{} & $\Theta_T$: MS-MARCO (RankingLlama) & .\textbf{099} & .\textbf{724} & .\textbf{770} & .\textbf{731} & .\textbf{265} & .\textbf{448} & .\textbf{373} & .\textbf{170} & .096 & .\textbf{182} & .188 & .129 & .860 & .\textbf{869} & .810 & .098 \\  

\multicolumn{1}{|c|}{} & $\Theta'$: Task Arithmetic ($\alpha=1$) & .096 & .718 & .757 &  .723  &  .265   &  .445   &  .370   &  .167  & .095  &  .179   &  .185   &  .126 & .858 & .867 & .801 & .098 \\ 

\multicolumn{1}{|c|}{} & $\Theta'$: Task Arithmetic (optimized $\alpha=0.8$) & .097 & .\textbf{728} & .765 &  .730  &  .262   &  .442   &  .365   &  .165  & .097  &  .\textbf{182}   &  .189   &  .129 & .\textbf{866} & .867 & .\textbf{812} & .\textbf{099}* \\ \midrule

\multicolumn{1}{|c|}{\multirow{4}{*}{\texttt{T5-Large}}}  & $\Theta_0$: Pre-trained  & .089 & .639 & .691 & .650 & .247 & .404 & .343 & .154 & .076 & .144 & .150 & .102 & .738 & .757 & .688 & .086 \\  

\multicolumn{1}{|c|}{} 
& $\Theta_D$: Domain-specific (SciFive)
& .091 & .637 & .691 & .649 & .247 & .404 & .343 & .154 & .065 & .111 & .121 & .082 & .596 & .582 & .551 & .076 \\  

\multicolumn{1}{|c|}{} & $\Theta_T$: MS-MARCO (Mono-T5) & .\textbf{095} & .\textbf{706} & .\textbf{743} & .\textbf{709} & .\textbf{266}* & .\textbf{431} & .\textbf{368}* & .\textbf{167} & .095 & .170 & .182 & .124 & .784 & 805 & .735 & .092 \\  


\multicolumn{1}{|c|}{} & $\Theta'$: Task Arithmetic ($\alpha=1$) & .092 & .688 & .721 &  .692  &  .257   &  .420   &  .356   &  .161  & .096  &  .\textbf{176}   &  .185   &  .124 & .816 & .\textbf{818} & .\textbf{765} & .096 \\ 

\multicolumn{1}{|c|}{} & $\Theta'$: Task Arithmetic ($\alpha=0.9$) & .092  & .699 & .727  &  .699   & .259 &  .423  &  .359   &  .162   &  .\textbf{098}*   &  .\textbf{176} &  .\textbf{187}*  &  .\textbf{126}   &  .\textbf{818}   &    .805 &  .759 & .\textbf{097}* \\ \midrule

\multicolumn{1}{|c|}{\multirow{4}{*}{\texttt{T5-base}}} & $\Theta_0$: Pre-trained  & .091 & .638 & .691 & .649 & .247 & .404 & .343 & .154 & .081 & .152 & .156 & .105 & .710 & .761 & .671 & .084 \\  

\multicolumn{1}{|c|}{} 
& $\Theta_D$: Domain-specific (SciFive)
& .090 & .638 & .691 & .648 & .248 & .400 & .343 & .154 & .059 & .110 & .115 & .080 & .646 & .662 & .598 & .079 \\ 

\multicolumn{1}{|c|}{} & $\Theta_T$: MS-MARCO (Mono-T5) & .096 & .681 & .726 & .684 & .258 & .\textbf{424} & .\textbf{359} & .\textbf{162} & .090 & .163 & .173 & .118 & .762 & .782 & .712 & .089 \\  

\multicolumn{1}{|c|}{} & $\Theta'$: Task Arithmetic ($\alpha=1$) & .089 & .645 & .686 &  .649  &  .250   &  .405   &  .345   &  .156  & .070  &  .131   &  .136   &  .094 & .764 & .825 & .726 & .088 \\ 

\multicolumn{1}{|c|}{} & $\Theta'$: Task Arithmetic ($\alpha=0.7$) & .\textbf{098} & .\textbf{702} & .\textbf{748} &  .\textbf{708}*  &  .\textbf{259}   &  .\textbf{424}   &  .358   &   .\textbf{162}  & .\textbf{091}  &  .\textbf{171}   &  .\textbf{176}   &  .\textbf{120} & .\textbf{798} & .\textbf{836} & .\textbf{753} & .\textbf{095}* \\ \midrule

\multicolumn{1}{|c|}{\multirow{4}{*}{\texttt{RoBERTa-base}}} & $\Theta_0$: Pre-trained  & .090 & .633 & .686 & .646 & .248 & .405 & .344 & .154 & .072 & .139 & .142 & .096 & .612 & .681 & .579 & .077 \\  

\multicolumn{1}{|c|}{} 
& $\Theta_D$: Domain-specific (BioMed RoBERTa) & .091 & .639 & .691 & .649 & .241 & .408 & .340 & .154 & .071 & .138 & .140 & .096 & .600 & .657 & .561 & .077 \\  

\multicolumn{1}{|c|}{} & $\Theta_T$: MS-MARCO (msmarco-RoBERTa) & .095 & .655 & .707 & .662 & .258 & .425 & .359 & .162 & .087 & .165 & .170 & .116  & .776 & .818 & .732 & .090 \\

\multicolumn{1}{|c|}{} & $\Theta'$: Task Arithmetic ($\alpha=1$) & .092 & .649 & .700 &  .659  &  .250   &  .408   &  .347   &  .156  & .078  &  .153   &  .156   &  .108 & .784 & .\textbf{822} & .734 & .089 \\ 

\multicolumn{1}{|c|}{} & $\Theta'$: Task Arithmetic ($\alpha=0.3$) & .\textbf{096} & .\textbf{669} & .\textbf{720} & .\textbf{676} &  .\textbf{259}  &  .\textbf{432}   &  .\textbf{361}   &  .\textbf{165}  & .\textbf{088} & .\textbf{169} &  .\textbf{173} & .\textbf{118}  & .\textbf{806}* & .821 & .\textbf{757}* & .\textbf{093}*  \\ \midrule

\multicolumn{1}{|c|}{\multirow{4}{*}{\texttt{DistilBERT}}} &  $\Theta_0$: Pre-trained  & .091 & .635 & .689 & .647 & .246 & .407 & .345 & .156 & .082 & .153 & .159 & .108 & .712 & .745 & .666 & .084 \\   

\multicolumn{1}{|c|}{} 
& $\Theta_D$: Domain-specific (Bio-DistilBert)
& .093 & .661 & .706 & .663 & .251 & .405 & .346 & .155 & .083 & .151 & .160 & .108 & .736 & .789 & .697 & .086 \\  

\multicolumn{1}{|c|}{} & $\Theta_T$: MS-MARCO (msmarco-distilbert) & .093 & .657 & .703 & .662 & .\textbf{258} & .418 & .357 & .161 & .087 & .159 &  .168 & .115  & .794 & .808 & .744 & .091 \\

\multicolumn{1}{|c|}{} & $\Theta'$: Task Arithmetic ($\alpha=1$) & .090 & .641 & .689 &  .650 &  .249  &  .406   &  .346   &  .156   &  .076  & .141  &  .147   &  .099    & .710 & .745 & .675 & .083 \\ 

\multicolumn{1}{|c|}{} & $\Theta'$: Task Arithmetic ($\alpha=0.5$) & .\textbf{095} & .\textbf{671} & .\textbf{720} & .\textbf{677} &  .257  &  .\textbf{429}   &  .\textbf{359}   &  .\textbf{163}  & .\textbf{088} & .\textbf{162} &  .\textbf{171} & .\textbf{116}  & .\textbf{806} & .\textbf{849} & .\textbf{765} & .\textbf{094}*  \\ 

 \bottomrule

\end{tabular}
}
\caption{Effectiveness of all models on Biomedical and Scientific domains. Best results are highlighted in boldface.}
\label{table_scientific}
\end{table*}

\subsection{Task Arithmetic for Zero-Shot IR}
\label{sec:tzir}

In this Section we
detail how Task Arithmetic 
can be applied in the context of
Information Retrieval, particularly for domain and language transfer in zero-shot conditions. 
Let \( \Theta_0 = \{(\theta_1)_0, \dots, (\theta_N)_0\} \) denote the parameters of a pre-trained LLM. Fine-tuning this model on a generic IR task (e.g., on the MS-MARCO benchmark \cite{MSMARCO}) produces \( \Theta_T = \{(\theta_1)_T, \dots, (\theta_N)_T\} \), while fine-tuning \(\Theta_0\) on a specific domain yields \( \Theta_D = \{(\theta_1)_D, \dots, (\theta_N)_D\} \).
We define the Task Vector \(\tau_D\) for domain \(D\) as follows:
\begin{equation}
\label{eq.taskvector}
\tau_D = \{\tau_1, \dots, \tau_N\}, \quad \text{where} \quad \tau_i = (\theta_i)_D - (\theta_i)_0.
\end{equation}
This vector \(\tau_D\) represents the domain-specific shift in the parameter space. 
To create a domain-aware IR model \(\Theta'\), we add \(\tau_D\) to the IR-tuned model \(\Theta_T\):
\begin{equation}
\label{eq.sumvector}
\Theta' = \{\theta_i' = (\theta_i)_T + \alpha \tau_i\}_{i=1}^{N}.
\end{equation}
The scaling factor \(\alpha\in \mathbb{R}\) controls how much of the domain vector is injected. If \(\alpha=0\), then \(\Theta'\) defaults to \(\Theta_T\). Setting \(\alpha>0\) adds the specialized knowledge, while \(\alpha<0\) subtracts 
it. In a fully zero-shot scenario, which does not require additional labelled data, $\alpha$ is equal to one. If, instead, a small development set is available, the hyperparameter \(\alpha\) can be optimized.

In summary, our framework involves three models: i) a publicly available pre-trained LLM, i.e. $\Theta_0$, ii) the $\Theta_0$ LLM fine-tuned on a specific domain, 
i.e. $\Theta_D$, and iii) the $\Theta_0$ LLM fine-tuned for IR, i.e. $\Theta_T$.  
In many publicly released models,
$\Theta_D$ is trained using a Language Modeling (LM) or Masked Language Modeling (MLM) objective, while $\Theta_T$ is fine-tuned to specific IR tasks like re-ranking or passage retrieval.  
As shown in Figure~\ref{fig:task-vector-img}, we follow a three-step procedure:
    (1) 
    \textbf{Task Vector Generation}: Using Equation~\ref{eq.taskvector}, we compute $\tau_D$ by subtracting the pre-trained model’s weights $\Theta_0$ from the domain-fine-tuned model’s weights $\Theta_D$. This step captures the domain shift in parameters.  
    (2) 
    \textbf{Task Vector Integration}: Following Equation~\ref{eq.sumvector}, we add the task vector $\tau_D$ (scaled by $\alpha$) to the IR-specific model $\Theta_T$, obtaining the $\Theta'$ model. This operation seamlessly transfers domain knowledge into the IR model without needing further backpropagation or labeled domain data.
    (3) 
    \textbf{Zero-Shot Evaluation}: Finally, we directly evaluate the adapted model $\Theta'$ on IR tasks in the target domain. As no training steps are required, we can readily test multiple domains, languages, or tasks by simply substituting different Task vectors.

\section{Experimental Setup}
\label{ES}

In this Section, we describe the datasets, evaluation metrics, and models used to assess the effectiveness of our proposed approach. The evaluation spans scientific, biomedical, and multilingual scenarios to show the potential of Task Arithmetic for zero-shot IR.

\subsection{Datasets and Evaluation Metrics}

We evaluate our approach on eight publicly available datasets across diverse domains. Four of these datasets are drawn from the BEIR benchmark \cite{thakur2021beir}: TREC-COVID \cite{Trec_covid2021} and NFCorpus \cite{boteva2016fullnfcorpus} address biomedical retrieval, while SCIDOCS \cite{cohan2020specterscidocs} and SciFact \cite{wadden2020scifact} focus on scientific citation prediction and fact-checking, respectively. The remaining four datasets concern language-specific retrieval: GermanQuAD \cite{moller2021germanquad}, which targets question answering in German, and three subsets (English, French, and Spanish) from the MIRACL multilingual IR challenge \cite{zhang2023miracl}. 
We focus on the biomedical and scientific domains, as they represent areas where both domain-specific and IR-focused models are available, thus enabling the application of Task Arithmetic. In contrast, we exclude BEIR datasets based on Wikipedia
since the pre-trained language models used in our experiments have already been trained on Wikipedia \cite{devlin2019bert,liu2019robertarobustlyoptimizedbert, raffel2020exploring, touvron2023llama2openfoundation}, and we cannot add domain-specific knowledge through Task Arithmetic.
Retrieval effectiveness is measured using P@10, NDCG@3, NDCG@10 and MAP@100. Statistical significance is assessed via a Bonferroni-corrected two-sided paired student’s t-test at $99\%$ confidence. In all result tables, the symbol * 
indicates a statistically significant improvement over the best performing baseline. 
\subsection{Models and Baselines}
\begin{table*}[htp]
\resizebox{\linewidth}{!}{
\begin{tabular}{c|c|cccc|cccc|cccc|cccc|}

\multicolumn{2}{c|}{\textbf{Model}} & \multicolumn{4}{c|}{\textbf{GermanQuAD}}   & \multicolumn{4}{c|}{\textbf{MIRACL Spanish}} & \multicolumn{4}{c|}{\textbf{MIRACL French}} & \multicolumn{4}{c}{\textbf{MIRACL English}} \\ \midrule

\multicolumn{1}{|c|}{\textbf{Re-ranker}} & Variant & \small{P@10} & \small{NDCG@3} & \small{\small{NDCG@10}} & \small{MAP@100}  & \small{P@10} & \small{NDCG@3} & \small{NDCG@10}  & \small{MAP@100}  & \small{P@10} & \small{NDCG@3} & \small{NDCG@10}  & \small{MAP@100} & \small{P@10} & \small{NDCG@3} & \small{NDCG@10}  & \small{MAP@100} \\ \midrule

\multicolumn{2}{|c|}{BM25} & .059 & .381 & .437 & .397  & .135 & .248 & .270 & .215 & .052  & .125 & .174 & .139 & .107 & .251 & .302 & .247  \\ \midrule

\multicolumn{1}{|c|}{\multirow{4}{*}{MT5-base}} & $\Theta_0$: Pre-trained & .050 & .275 & .336 & .296 &  .094 & .173 & .191 & .152 & .035 & .095 & .127 & .106 & .075 & .175 & .212 & .174  \\    

\multicolumn{1}{|c|}{} 
& $\Theta_D$: Language specific & .051 & .296 & .353 & .316 & .097 & .172 & .191 & .153 & .045 & .099 & .143 & .113  & .079 & .180 & .225 & .187 \\ 

\multicolumn{1}{|c|}{} & $\Theta_T$: MS-MARCO (en) & .069 & .451 & .513  & .463  & .190 & .342 & .379 & .301 & .071 & .175 & .234 & .186 & .140 & .330 & .398 & .325  \\  

\multicolumn{1}{|c|}{} & $\Theta'$: Task Arithmetic ($\alpha=1$) & .\textbf{071}* &  .\textbf{477}*   &  .\textbf{537}*   &  \textbf{487}*   &   .\textbf{200}*  &  .\textbf{373}*   &  .\textbf{405}*   &  .\textbf{325}*   &  .\textbf{081}*   &  .\textbf{215}*  &  .\textbf{278}*   &  .\textbf{220}* & \textbf{.151}* & \textbf{.366}* & \textbf{435}* & \textbf{.358}* 


\\ \midrule

\end{tabular}
}
\caption{Effectiveness of all models on Language Transfer. Best results are highlighted in boldface.}
\label{table_multi_lingua}
\end{table*}
\begin{sloppypar}
For our experiments, we focus on two-stage retrieval and we use six different pre-trained language models (i.e. $\Theta_0$) spanning multiple retrieval paradigms (bi-encoder, cross-encoder, LLM) and neural architectures (encoder-only, encoder-decoder, decoder-only). In details, we use \texttt{DistilBERT} \cite{sanh2019distilbert} and \texttt{RoBERTa-base} \cite{liu2019robertarobustlyoptimizedbert} as encoder-only bi-encoders, \texttt{T5-base}, \texttt{T5-Large} \cite{raffel2020exploring}, and \texttt{MT5-base} \cite{xue2021mt5massivelymultilingualpretrained} as encoder-decoder cross-encoders, and \texttt{LLama-2-7b} \cite{touvron2023llama2openfoundation} as a decoder-only LLM. 
For each pre-trained model, we compute Task Vectors by subtracting the weights of a publicly available domain- or language-specific fine-tuned model from those of its original pre-trained version. Specifically, we use \texttt{LLama2-MedTuned-7b} \cite{rohanian2024exploring}, \texttt{SciFive} \cite{phan2021scifive}, \texttt{BioMed-RoBERTa} \cite{biomedRoberta}, and \texttt{Bio-DistilBERT} \cite{rohanian2023effectiveness} for the biomedical and scientific domains, as well as \texttt{MT5-base-german}, \texttt{MT5-base-spanish}, \texttt{MT5-base-french}, and \texttt{MT5-base-english} \cite{calizzano-etal-2022-generating} for language adaptations. These Task Vectors are then added to models fine-tuned on MS-MARCO (\texttt{RankingGPT-Llama2-7b} \cite{zhang2023rankinggpt}, \texttt{MonoT5} \cite{nogueira-etal-2020-document}, \texttt{msmarco-RoBERTa} \cite{reimers-2019-sentence-bert}, \texttt{msmarco-distilbert} \cite{reimers-2019-sentence-bert}, and \texttt{MT5-base-msmarco} \cite{bonifacio2021mmarco}), a widely adopted passage retrieval benchmark \cite{MSMARCO}. 
This setup facilitates direct comparisons with the same models specialized for either a specific language or domain (i.e. $\Theta_D$), or IR (i.e. $\Theta_T$).
In a fully zero-shot scenario, we set $\alpha=1$. Furthermore, in a setting where few labeled data are available,
we tune the scaling factor $\alpha$ from 0.1 to 1.0 in steps of 0.1, selecting the optimal value based on the highest average retrieval performance over two development sets: the official NFCorpus split and a 20\% subset of SciFact training queries. Since GermanQuAD and MIRACL do not provide development sets, we apply 
a fully zero-shot scenario by not optimizing the value of $\alpha$, i.e. we put $\alpha=1$.
We report both the results with the optimized and not-optimized $\alpha$.
All re-ranking experiments begin by retrieving the top 100 documents via BM25. Following a two-stage retrieval paradigm, the final rankings are then computed using a weighted sum of BM25 and LLM scores, with $\lambda_{BM25}$ and $\lambda_{LLM}$ optimized in $[0,1]$ on the NFCorpus and SciFact development sets. We take the average score for all remaining datasets, i.e. $\lambda_{BM25}=\lambda_{LLM}=0.5$.
\end{sloppypar}

\section{Results and Discussion}
\label{Results}

Table~\ref{table_scientific} presents the performance of our approach on the biomedical and scientific datasets, evaluated with five different models. 
In the initial evaluation, we fix $\alpha=1$. Under this setting, Task Arithmetic outperforms the MS-MARCO fine-tuned baselines only on TREC-COVID with RoBERTa-base, T5-base, and T5-Large, and on SCIDOCS with T5-Large. 
These findings suggest the need for a small amount of labeled data to optimize the value of $\alpha$. The remainder of this section focuses on the results obtained when $\alpha$ is optimized.
Across all datasets and metrics, the Task Arithmetic–based model (\(\Theta'\)) 
consistently outperforms BM25, indicating its potential for effective domain adaptation in IR. 
For bi-encoders (DistilBERT and RoBERTa-base), Task Arithmetic yields consistent improvements over all baselines, with one exception: DistilBERT on the NFCorpus shows a minor drop in P@10 compared to the MS-MARCO fine-tuned counterpart (\(\Theta_T\)). Nevertheless, our approach achieves a statistically significant improvement in MAP@100 on TREC-COVID, surpassing every baseline. 
For cross-encoders (T5-base and T5-Large), Task Arithmetic outperforms MonoT5 (\(\Theta_T\)) on SCIDOCS and TREC-COVID, while producing comparable results on SciFact and NFCorpus. Notably, our method shows significant gains in MAP@100 on TREC-COVID, on the T5 variant, and yields statistically significant improvements in NDCG@10 on TREC-COVID and SCIDOCS. 
Regarding the decoder-only model (LLama-2), our approach (\(\Theta'\)) achieves superior performance on TREC-COVID compared to all baselines and remains competitive on SCIDOCS. Interestingly, the pre-trained LLama-2 (\(\Theta_0\)) attains the highest P@10 and NDCG@10 on SciFact and SCIDOCS, which likely reflects the extensive domain knowledge acquired during large-scale pre-training \cite{touvron2023llama2openfoundation}. Finally, it is worth noting that the optimal scaling factor \(\alpha\) exceeds 0.3 for all models and surpasses 0.7 for T5 variants and LLama-2, indicating that Task Arithmetic injects non-trivial domain knowledge into these retrieval models. 

Table~\ref{table_multi_lingua} extends the results to multilingual IR by using MT5-base as a cross-encoder for language-specific tasks. Both our 
proposed approach (\(\Theta'\)) and the IR-specific baseline (\(\Theta_T\)) outperform all other baselines on each metric. Notably, our method shows statistically significant improvements over the IR-specific model by up to 18\% in NDCG@10, highlighting its effectiveness in adapting to new language settings. The pre-trained MT5-base (\(\Theta_0\)) and its language-specific variant (\(\Theta_D\)) do not outperform BM25, suggesting that these models are not inherently optimized for IR tasks. In contrast, our approach successfully injects language-specific knowledge into the multilingual IR model (\(\Theta_T\)), thereby enhancing its retrieval capabilities. 
These results further support Task Arithmetic as a lightweight but powerful strategy for zero-shot adaptation in multilingual IR.

\section{Ablation Study}
\label{Ablation}
We conduct an ablation study to examine scaling factor 
impact while applying
Task Arithmetic. 
Table~\ref{table_lambda_orizz} presents NDCG@10 scores on NFCorpus and SciFact development sets for $\alpha$ values from 0.1 to 1.0 in increments of 0.1. 
The results reveal that retrieval performance varies considerably with changes in $\alpha$. On SciFact, for instance, T5-base improves from .640 at $\alpha=1.0$ to .722 at $\alpha=0.7$, representing a notable difference of approximately 12\%. In general, while values in the 0.7–0.9 range often yield strong results for 
models such as T5-base and T5-Large, 
no single $\alpha$ value is consistently optimal across all models or datasets. This variability underscores the importance of model-specific calibration.
Moreover, $\alpha=1.0$ rarely provides the best performance, suggesting that excessive emphasis on domain parameters can overshadow the IR-specific knowledge embedded in IR-tuned weights. RoBERTa-base and DistilBERT sometimes peak at moderate values between 0.3 and 0.5, whereas T5-based models and LLama-2 commonly favor higher settings. These observations indicate that a small grid search over $\alpha$ can be effective in identifying a 
tradeoff between the IR 
and the domain-specific knowledge. 
\begin{table}[t]
\centering
\resizebox{\linewidth}{!}{
\begin{tabular}{cccccccccccc}
\toprule
\textbf{Model} & \textbf{Dataset} & 0.1 & 0.2 & 0.3 & 0.4 & 0.5 & 0.6 & 0.7 & 0.8 & 0.9 & 1  \\ \midrule

\multirow{2}{*}{LLama 2-7B} & \textbf{SciFact} & .631 & .643 & .660 & .675 & .687 & .705 & .\textbf{712} & .711 & .705 & .704 \\
 & \textbf{NFCorpus} & .235 & .238 & .239 & .241 & .\textbf{242} & .241 & .239 & .\textbf{242} & .241 & .241 \\ \midrule

\multirow{2}{*}{T5-Large} & \textbf{SciFact} & .728 & .732 & .727 & .731 & .730 & .732 & .728 & .730 & .\textbf{737} & .710 \\
 & \textbf{NFCorpus} & .307 & .307 & .307 & .308 & .308 & .306 & .306 & .\textbf{311} & .\textbf{311} & .309 \\ \midrule
 
\multirow{2}{*}{T5-base} & \textbf{SciFact} & .712 & .710 & .712 & .711 & .709 & .718 & .\textbf{722} & .708 & .673 & .640 \\
 & \textbf{NFCorpus} & .302 & .301 & .304 & .306 & .306 & .311 & .\textbf{313} & .312 & .303 & .298 \\ \midrule

\multirow{2}{*}{RoBERTa-base} & \textbf{SciFact} & .713 & .717 & .\textbf{725} & .723 & .722 & .710 & .715 & .706 & .695 & .687 \\
 & \textbf{NFCorpus} & .334 & .331 & .332 & .332 & .333 & .\textbf{334} & .330 & .323 & .315 & .307 \\ \bottomrule

\multirow{2}{*}{DistilBERT} & \textbf{SciFact} & .723 & .720 & .723 & .722 & .\textbf{724} & .712 & .705 & .699 & 
.688 & .652 \\
 & \textbf{NFCorpus} & .334 & .\textbf{336} & .333 & .333 & .333 & .329 & .325 & .306 & .297 & .286 \\ \bottomrule
\end{tabular}
}
\caption{Ablation study about the impact of the scaling factor.} 
\label{table_lambda_orizz}
\end{table}

\section{Conclusion} 
\label{Conclusion}

In this paper, we investigate Task Arithmetic as a training-free method for zero-shot domain and language adaptation in IR, leveraging publicly available domain- and IR-specific LLMs. To this aim, 
we evaluate Task Arithmetic with six LLMs, including encoder-only, encoder-decoder, and decoder-only architectures,     
across scientific, biomedical, and multilingual datasets. Our analysis shows that the proposed approach consistently improves the IR-specific model's performance across the board, reaching gains of up to $18\%$ in NDCG@10. 
These findings underscore Task Arithmetic as a lightweight yet powerful strategy for IR applications, particularly when computational resources or labeled data are limited.



\section*{Acknowledgment}
We acknowledge the CINECA award under the ISCRA initiative, for the availability of high-performance computing resources and support. This work was partially supported by the European Union – Next Generation EU within the project NRPP M4C2, Investment 1.,3 DD. 341 - 15 march 2022 – FAIR – Future Artificial Intelligence Research – Spoke 4 - PE00000013 - D53C22002380006, and by the MUR under the grant “Dipartimenti di Eccellenza 2023-2027” of the Department of Informatics, Systems and Communication of the University of Milano-Bicocca, Italy.

\bibliographystyle{ACM-Reference-Format}
\bibliography{biblio}


\end{document}